\documentclass[reprint,aps,pre,showpacs,superscriptaddress,groupedaddress]{revtex4-1} 
\usepackage{graphicx,floatrow,subfig,amsmath,amssymb,dcolumn,bm,chngcntr}
\usepackage[font=small,format=plain,labelfont=bf,justification=justified,singlelinecheck=false]{caption}
\usepackage{subeqnarray}
\usepackage{color}

\def\p{\partial}
\def\be{\begin{eqnarray}}
\def\ee{\end{eqnarray}}
\def\bes{\begin{subeqnarray}}
\def\ees{\end{subeqnarray}}
\def\f{\frac}
\def\lp{\left(}
\def\rp{\right)}

\def\befi{\begin{figure}}
\def\eefi{\end{figure}}

\def\bce{\begin{center}}
\def\ece{\end{center}}

\def\e{{\rm e}}

\def\p{\partial}
\def\be{\begin{eqnarray}}
\def\ee{\end{eqnarray}}
\def\f{\frac}
\def\lp{\left(}
\def\rp{\right)}

\def\befi{\begin{figure}}
\def\eefi{\end{figure}}

\def\bce{\begin{center}}
\def\ece{\end{center}} 
\def\nn{\nonumber}

\catcode`\!=\active
\def!#1!{}
\def\ba#1#2\ea{\begin{align}\label{#1}#2\end{align}}
\def\bsa#1#2\esa{\begin{subequations}\label{#1} \begin{align}#2\end{align} \end{subequations}}

\begin{document}
\preprint{APS/123-QED}
\title{Predictability Horizon of Oceanic Rogue Waves}
\author{Reza Alam}
\affiliation{Department of Mechanical Engineering, University of California, Berkeley, CA, 94720}

\begin{abstract}
Prediction is a central goal and a yet-unresolved challenge in the investigation of oceanic rogue waves. 
Here we define a horizon of predictability for oceanic rogue waves and derive, via extensive computational experiments, the first statistically-converged predictability time-scale for these structures. 
We show that this time-scale is a function of the sea state as well as the strength (i.e. overall height) of the expected rogue wave. 
The presented predictability time-scale establishes a quantitative metric on the combined temporal effect of the variety of mechanisms that together lead to the formation of a rogue wave, and is crucial for the assessment of validity of rogue waves predictions, as well as for the critical evaluation of results from the widely-used model equations. The methodology and presented results can have similar implications in other systems admitting rogue waves, e.g. nonlinear optics and plasma physics.
\end{abstract}
\maketitle
\counterwithout{equation}{section}
\allowdisplaybreaks

Oceanic rogue waves are short-lived very large amplitude waves (a giant crest typically followed or preceded by a deep trough) that appear and  disappear suddenly in the ocean causing damages to ships and offshore structures\cite{Dysthe2008,Onorato2001}. They are rare-enough phenomena that to-date very few measured cases have been documented\cite{Mori2002}, but at the same time frequent enough that cause annually several incidents of damage to ships and offshore structures\cite{Draper1964,Draper1971,Bruun1994}. What mechanism(s) leads to formation of rogue waves is yet a matter of dispute, although decades of research have shed a lot of light on several aspect of their existence\cite{Kharif2003}. For instance, it is clear today that linear theories significantly under-estimate the frequency of occurrence of such waves\cite{Xiao2013} and nonlinear interactions and processes play a pivotal role in the series of events leading to formation of oceanic rogue waves\cite{Janssen2003}. 

The central question in the investigation of rogue waves is if they can be predicted; i.e., when and where they will occur, and what their features are. We are today closer than ever to answer this question owing to the advancement in the radar technology as well as surface reconstruction techniques: the state of the art radar technology can now provide accurate wave height measurement over large spatial domains\cite{Barale2008,Young1985,Dankert2004} and when combined with advanced wave-field reconstruction techniques\cite{Wu2004,Blondel2010} together render deterministic details of the current state of the ocean (i.e. surface elevation and velocity field) at any given moment of the time with a very high accuracy. This knowledge of the ocean state, that although small but has an inevitable uncertainty from both measurements and reconstruction, is known to be good enough for the prediction of average state of the ocean in the future
\footnote{The general problem of prediction of future of the ocean state is a classic challenge\cite{Massel1996,Golding1983}. Available results are, however, mostly either based on linear theory\cite{Zhang1999,Zhang1999a,Edgar2000,Abusedra2011} or phase-averaged models \cite{wam88,Booij1996} that cannot obtain deterministic details of oceanic wave fields necessary for prediction of extreme events such as rogue waves\cite{Pelinovsky2008}.}. 
%
But the important question remains whether with this knowledge the forthcoming oceanic rogue waves can be predicted and if so, how much in advance and with what accuracy?

Here, we define a horizon of predictability for oceanic rogue waves and establish, via an extensive statistical analysis, a predictability time-scale as a function of measurement uncertainties. This time-scale of predictability provides the first quantitative metric to evaluate the range of validity of prediction efforts, and, since it is obtained through primitive equations (i.e. Euler's equation) can set a standard in evaluating classical model equations (e.g. Nonlinear Schr\"{o}dinger equation \cite{Zakharov2013},{\footnote{For instance, it is known that the wave spectra associated with Peregrine soliton (as well as other rational solutions of the NLS) have a specific signature that appears at an early stage of evolution of such waves\cite{Akhmediev2011,Akhmediev2011a}.}}).

We consider propagation of waves on the surface of a homogeneous, incompressible and inviscid ocean of constant depth $h$. Let's define a Cartesian coordinate system with the $x,y$-axis on the mean free surface and $z$-axis positive upward. Assuming the flow is irrotational, a velocity potential $\phi$ can be defined such that $\bf u=\nabla \phi$ where $\bf u$ is the velocity vector in the fluid domain. The governing equation (conservation of mass), and boundary conditions (momentum equation and kinematic conditions) read
\vspace{-0.3cm}
\bsa{110}
&\nabla^2\phi=0,\hspace{3.5cm}-h<z<\eta({\bf x},t)\\
\nn&\phi_{tt}+g\phi_z+&\\
&~\hspace{0.3cm}[\p_t+1/2(\nabla\phi\cdot\nabla)](\nabla \phi\cdot\nabla\phi)=0,~~ z=\eta({\bf x},t)\\
&\phi_z=0,\hspace{5.0cm} z=-h. 
\esa
where $\eta({\bf x},t)=-[\phi_t+1/2(\nabla\phi.\nabla\phi)]/g$ is the surface elevation and $g$ is the gravity acceleration.

Consider, on the ocean surface, a broadband spectrum of propagating waves whose spectral density function, $S(\omega)$, is given by a JONSWAP spectrum (Joint North Sea Wave Project) in the form\cite{Hasselmann1973}

\vspace{-2mm}
\ba{400}
S(\omega)=\f{\alpha_p g^2}{\omega^5} \e^\beta\gamma^\delta
\ea
\vspace{-3mm}\\
where $\beta= -1.25\lp {\omega_p}/{\omega}\rp^4$ with $\omega_p$ being the spectrum's peak frequency,  $\alpha_p=H_s^2\omega_p^2/(16I_0(\gamma)g^2)$ with $H_s$ being significant wave height defined as four times the standard deviation of the surface elevation, $\delta=\exp[-(\omega-\omega_p)^2/(2\omega_p^2\sigma^2)]$, and $\sigma$=0.07,0.09 for respectively $\omega\le\omega_p$ and $\omega>\omega_p$. The peak enhancement factor $\gamma$ typically ranges between $1<\gamma<9$, and we choose, as is typical\cite{Hasselmann1973,Carter1982,Ochi2005}, a mean value of $\gamma=$3.3. Zeroth order moment $I_0(\gamma)$ varies in the range $0.2<I_0(\gamma)<0.5$ and is calculated numerically\cite{Carter1982b}; for our application $I_0(3.3)$=0.3. 

For the direct simulation of evolution of a wave-field initiated by the spectrum \eqref{400}, we utilize a phase-resolved high-order spectral technique\cite{Dommermuth1987,west1987} formulated based on Zakharov's equation\cite{Zakharov1968} that can take into account a large number of wave modes (typically N=$O(1000)$) and a high order of nonlinearity (typically M=$O(10)$) in the perturbation expansion in terms of the wave steepness
\footnote{The High-order spectral method is a numerical integration of potential Euler equation for a phase-resolved direct simulation of a nonlinear wave field. The scheme belongs to the category of Boundary Element Methods, but in order to be efficient relies on spectral expansion. High-order spectral method was first developed for the problem of nonlinear wave-wave interaction in a homogeneous water \cite{Dommermuth1987,west1987}, but then extended to include seabed variations \cite{Liu1998,Alam2010}, density stratification \cite{alam2009b,Alam2012a,Alam2012b,Alam2009c}, and compliancy of the seabed (for the wave-mud interaction problem) \cite{Alam2011b,Alam2012d}. The scheme has been successfully cross validated with laboratory experiments as well as other (more restrictive) numerical results such as those from broadband modified nonlinear Schrodinger equation\cite{Toffoli2010}.}
. The scheme has already undergone extensive convergence tests as well as validations against experimental and other numerical results \cite{Liu1998,alam2009b,Alam2009c,Alam2010,Toffoli2010,Alam2012a,Alam2012c}. 

To quantify the effect of uncertainty on the predictability of oceanic rogue waves a three step procedure is followed: 1- We find an initial sea state\footnote{Sea State is a measure of how energetic the sea surface is. Based on Douglas Sea Scale, which is one of the most widely used scales, sea states range from 0 (calm) to 9 (phenomenal). Each sea state corresponds to a range of significant wave heights\cite{Metoffice2010}.} 
that after a specific time $t=t_r$ develops a rogue wave near $x=x_r$, 2- To include the effect of uncertainty in the initial condition, random  perturbations with a Gaussian distribution (that has a zero mean such that the overall energy of the spectrum stays the same) are added to both amplitude and phases of the initial state of the ocean, 3- The new perturbed initial condition is evolved and vicinity of $t=t_r,x=x_r$ is searched for rogue (or large) waves. This wave is compared with the rogue wave that the unperturbed system foresees. For each initial condition, steps one to three are repeated for a large set of initial perturbations until converged averaged quantities are obtained.

Finding an initial broadband sea state that leads to a rogue wave at a specific moment in the future is, however, a challenge. This challenge is  further highlighted in a statistical investigation where a large number of such cases are needed. To overcome this issue, we propose a technique that relies on reversibility of nonlinear governing equations of oceanic waves. Specifically, if $(\eta,\phi)$ is a solution to governing equations \eqref{110} in forward time, then $(\eta,-\phi)$ is a solution to the same set of equations in the reverse time, i.e., when $t$ is replaced by $-t$. In a forward-time simulation of governing equation \eqref{110}, if a rogue wave is observed at the time $t=t_i$ then we continue the simulation up to a final time $t_f=t_i+t_r$. At $t=t_f$ water surface elevation and potential, i.e. $\eta(\textbf x,t_f)$  and $\phi(\textbf x,t_f)$, are recorded. A direct simulation with initial surface elevation $\eta_0\equiv\eta(\textbf x,t_f)$ and initial potential $\phi_0\equiv-\phi(\textbf x,t_f)$ will result in a rogue wave at exactly $t=t_r$%
\footnote{It is to be noted that water waves show instability as a result of a number of nonlinear mechanisms such as the well-studied Benjamin-Feir instability, or, myriad of wave-wave resonances. These instabilities initially predict a one-way (i.e. non-reversible) energy transfer. A longer-time theoretical analysis (e.g. by a multiple-scales approach\cite{Mei1985}) which is now backed by experimental proof \cite{VanSimaeys2002}, however, reveals that after a threshold the process reverses and eventually the initial state is recovered. 

Similar recursion is also known to exist, and has validated experimentally, in the case of Peregrine solitons \cite{Peregrine1983,Shrira2009,Akhmediev2011a}. In the context of oceanic waves, under the assumption of weakly nonlinear monochromatic waves in deep water, the slow evolution of the wave envelope is governed by the Nonlinear Schrodinger Equation. Some exact solutions to the NLS have characteristics that match physical observations of rogue waves. For example Peregrine soliton which is a closed--form unsteady solution (algebraic breather) to the NLS \cite{Peregrine1983,Dysthe1999}, Akhmediev breather which is a space periodic\cite{akhmediev1986,Osborne2014,VishnuPriya2013,Mahnke2012}, and Kuznetsov-Ma breather which is a time-periodic solution to the NLS\cite{Kuznetsov1977,Ma1979,Zakharov2013}. 

Therefore, the assumption of reversibility of water waves is not necessarily violated by the one-way initial exponential growth of perturbations. Computational results presented in this paper consistently endorse this fact by showing excellent agreement in the reverse simulation.}.

For implementation of this procedure in a specific (given) sea state, we initialize our phase-resolved spectral scheme with amplitudes and frequencies given by the spectrum \eqref{400}, and with random phases that have a uniform distribution. Via running the direct computation for a relatively short initial time $t_0$ (typically $t_0 \sim O(20 T_p)$ where $T_p=2\pi/\omega_p$ is the period of the peak frequency wave in the spectrum), we search for those initial phases that lead to a rogue wave event within $0<t_i<t_0$. We found that the search for rogue waves is more efficient if $t_0$ is chosen smaller and instead more tests are conducted. We have performed $O(10^4)$ initial runs for each of the sea states four, five and six (respectively moderate, rough and very rough seas with $H_s$=1.875, 3.25, 5.00 meters, and $T_p$=8.8, 9.7, 12.4 sec) and for each sea state have collected a database of $O(100)$ cases from those initial conditions that lead to rogue waves. Water surface of the sea state five at the time of occurrence of a rogue wave with $H_{rs}=H_r/H_s$=2.78 ($H_r$ being the crest to trough height of the rogue wave) is shown in figure 1a. Our computational experiments show lower sea states do not develop rogue waves with very large values of $H_{rs}$. Specifically, in our database for the sea state four, Max($H_{rs}$)$\sim$2.6, whereas for sea states five and six we have recorded several rogue waves with $H_{rs}>$2.6 (c.f. fig. 3).

\begin{figure}
\includegraphics[width=8.0cm]{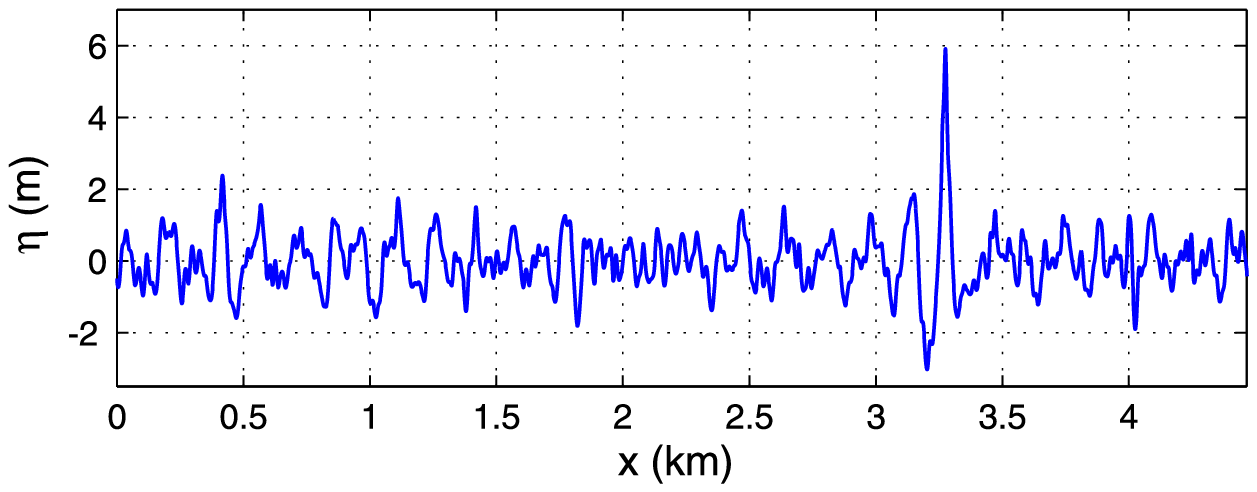}
\put(-15.0,75){(a)}\\
~~\includegraphics[width=8.1cm]{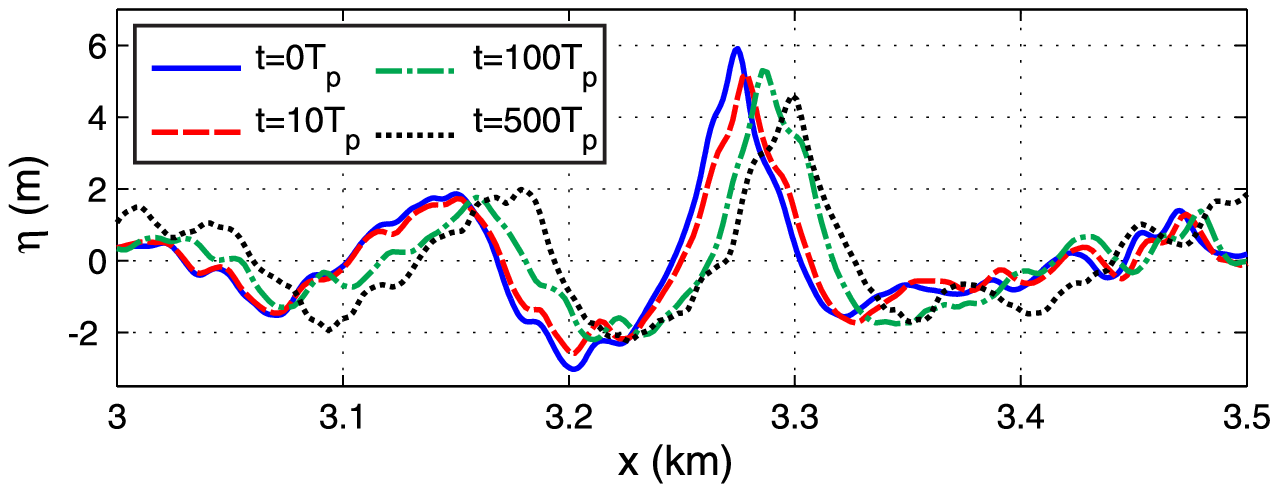}
\put(-18.0,75){(b)}
\caption{a. A rogue wave with $H_{rs}$=2.78 in the sea state five. b. Result of a $E_r$=10\% error in the estimation of the initial state of the sea on the prediction of rogue wave. If sea state is known at $t_r$=10$T_p$ before the occurence of the rogue wave then a $E_r$=10\% error has resulted in 13\% error in the predictions (red dashed line). For $t_r$=100$T_p$ and 500$T_p$ in advance (green dash-dotted line and black dotted line) the same initial uncertainty results in respectively 30\% and 28\% error. In the latter two cases the predicted wave is hardly a rogue wave by definition ($H_{ps}$=1.95,2.01). Simulations parameters are $N=$2048, $M$=4 and $T_p/\delta t$=128 for which presented results are converged.}
\end{figure}

If an initial search is successful and a rogue wave is obtained at $t_i<t_0$, we then continue the evolution up to $t_f=t_i+t_r$ where $t_r=nT_p$ with $n\sim O(10^{2-3})$. As discussed above, a new simulation with initial conditions ($\eta_0,\phi_0$)=($\eta(\textbf x,t_f)$,$-\phi(\textbf x,t_f)$) will lead to a rogue wave exactly at $t=t_r$. In practice, this initial condition is an outcome of combined measurements and reconstruction procedures\footnote{A proper number and distribution of wave probes can guarantee uniqueness of reconstruction of the wavefield (c.f. \cite{Wu2004}, \S 3.3.5, Equation (3.13) that provides a necessary and sufficient condition on measurements for the reconstruction to be unique)}, and is available with an inevitable \textit{range of uncertainty}. To quantify the effect of uncertainty on the predictability of a rogue wave, we add Gaussian perturbations with a zero mean and standard deviation $E_r$ (in percent) to the amplitude and phases of components of the initial condition ($\eta_0,\phi_0$). The perturbed state of the ocean, i.e. $\eta_{0p},\phi_{0p}$, is then used as the initial condition and its prediction at $t=t_r$ is compared with the actual rogue wave. In practice, if the perturbed initial condition predicts a \textit{close-enough} approximation of the height, location and the time of occurrence of the rogue wave most of the prediction objective is met. To take this fact into account, we search for the highest wave in the time span of $t_r-T_p<t<t_r+T_p$ and it is further checked that this highest wave is in the $\pm \lambda_p$ vicinity of the expected rogue wave. This highest  predicted wave (with a trough to crest height of $H_p$) is what our predictor foresees at the vicinity of where the actual rogue wave will occur. We define $H_{ps}\equiv H_p/H_{s'}$ with $H_{s'}$ is the significant waveheight as is calculated by the predictive simulation at the time of occurrence of $H_p$. Clearly if $E_r$=0, then $H_{ps}=H_{rs}$.

Effect of a $E_r$=10\% uncertainty in the initial condition on the prediction of the rogue wave of figure 1a is shown in figure 1b. At $t=0$ original rogue wave is obtained, but if the required prediction time $t_r$ is longer, the effect of the initial uncertainty is more highlighted. Specifically for $t_r=10T_p, 100T_p$ and $500T_p$, we have respectively $H_{ps}$=2.41, 1.95 and 2.01 corresponding to 13\%, 30\% and 28\% error. In fact for the latter tow cases ($t_r$=100$T_p$, 500$T_p$) predicted wave is hardly considered a rogue wave by the definition. Note that figure 1b shows the effect of \textit{one} specific set of initial perturbations on the shape of the predicted rogue wave.

\begin{figure}
\centering ~ \\ \vspace{0.3cm}
\includegraphics[width=4.3cm,height=3.2cm]{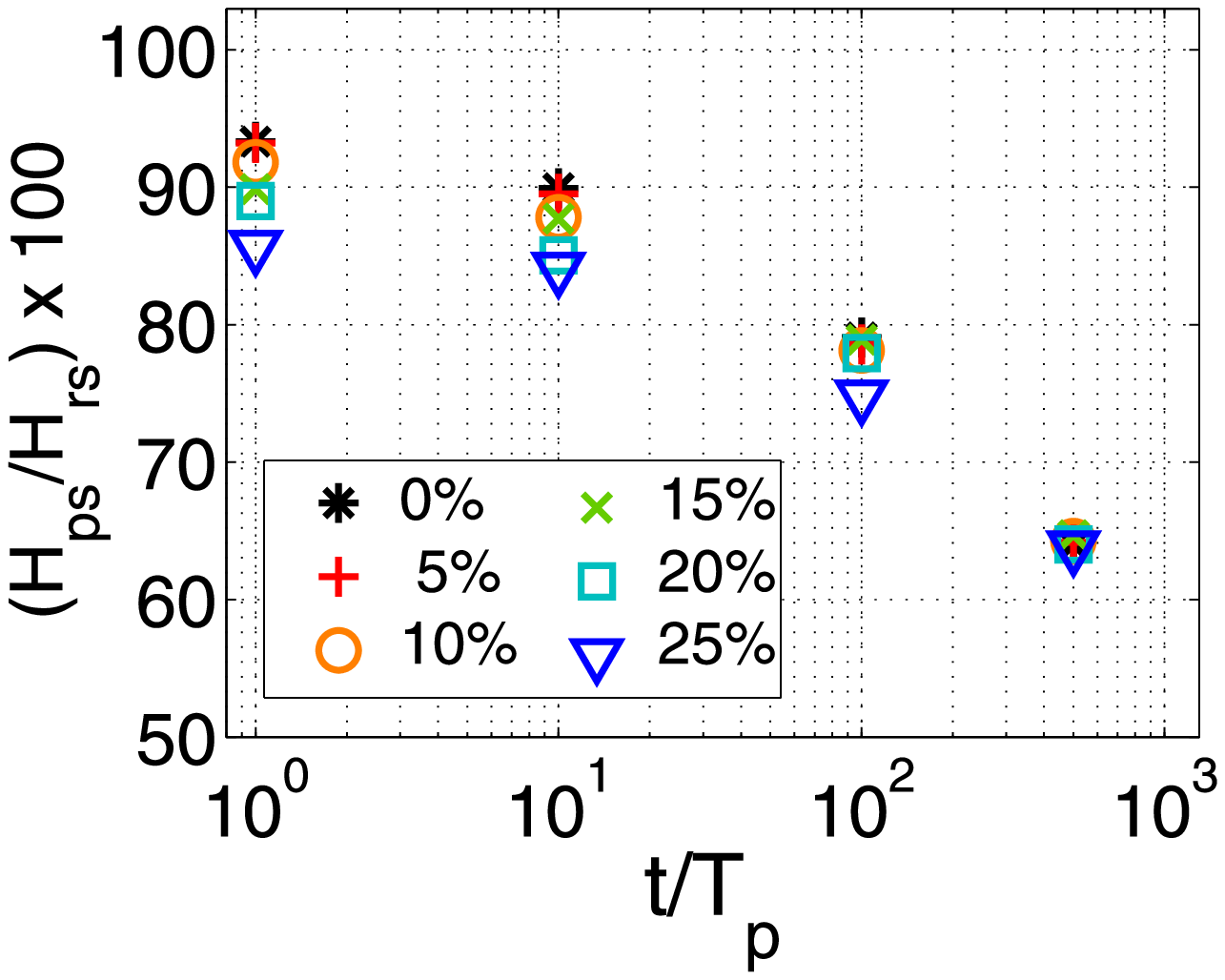}
\includegraphics[width=3.95cm,height=3.2cm]{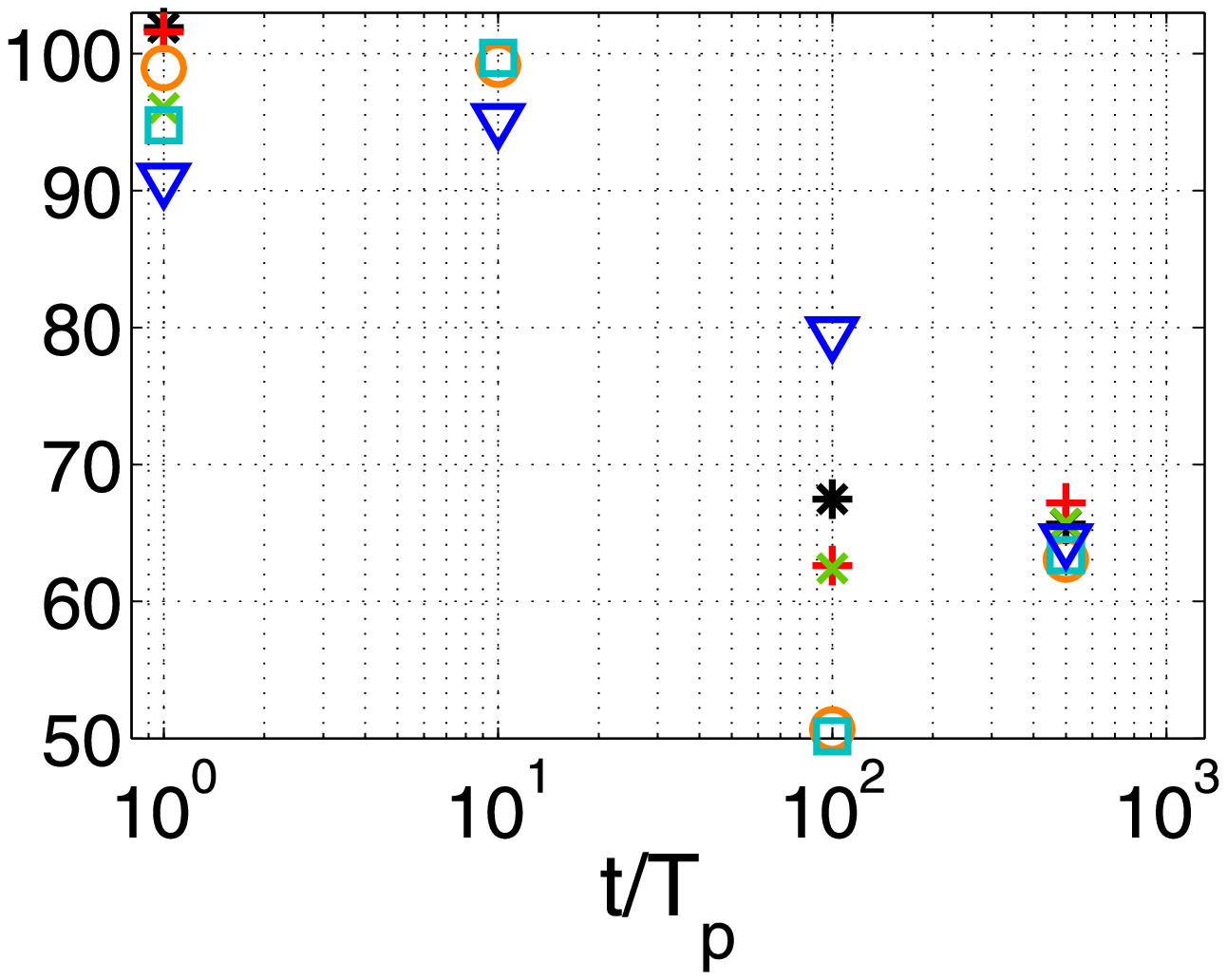}
\put(-130.0,27){(a)}    \put(-15.0,27){(b)}    \\
\includegraphics[width=4.3cm,height=3.2cm]{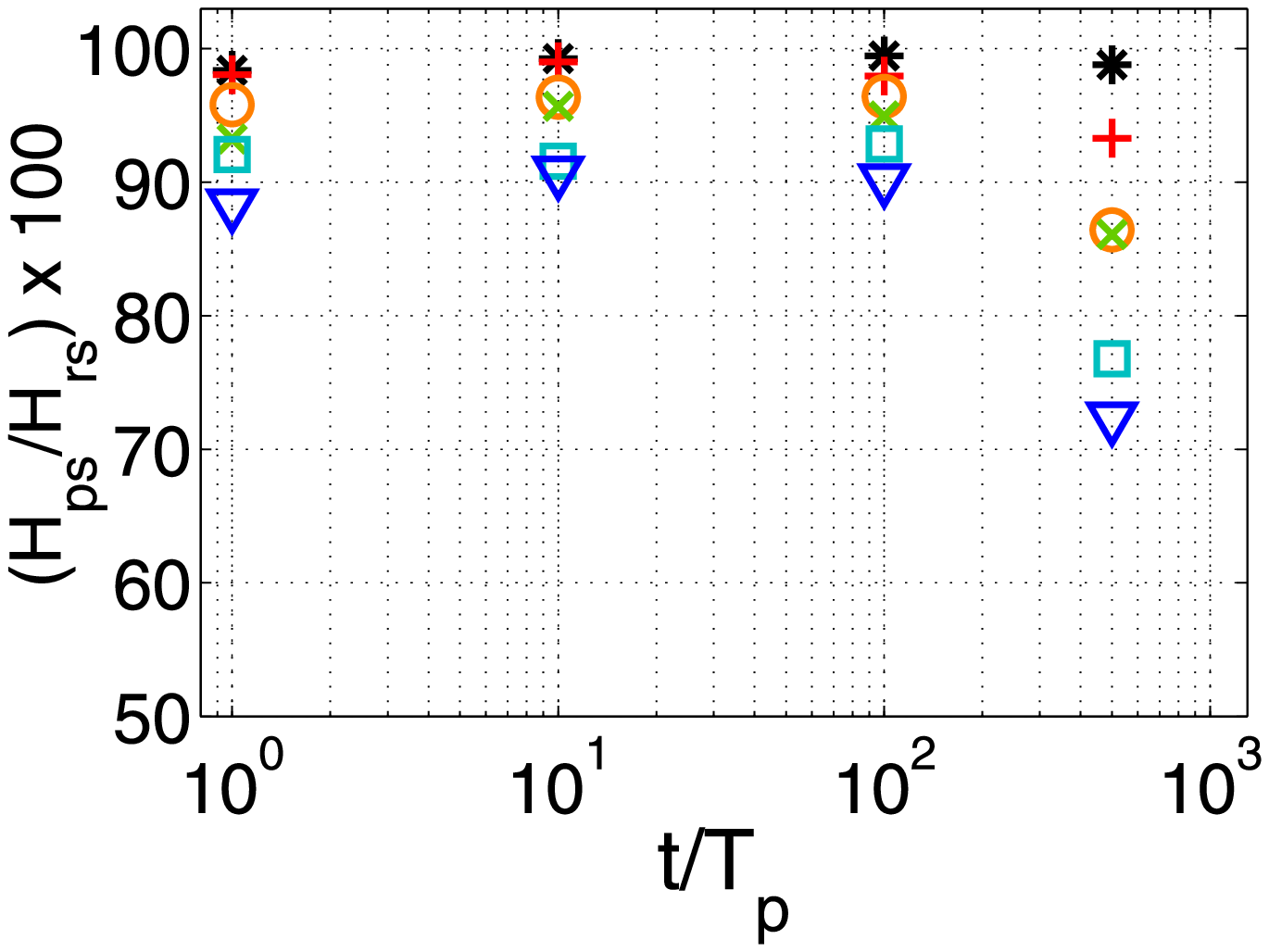}
\includegraphics[width=3.95cm,height=3.2cm]{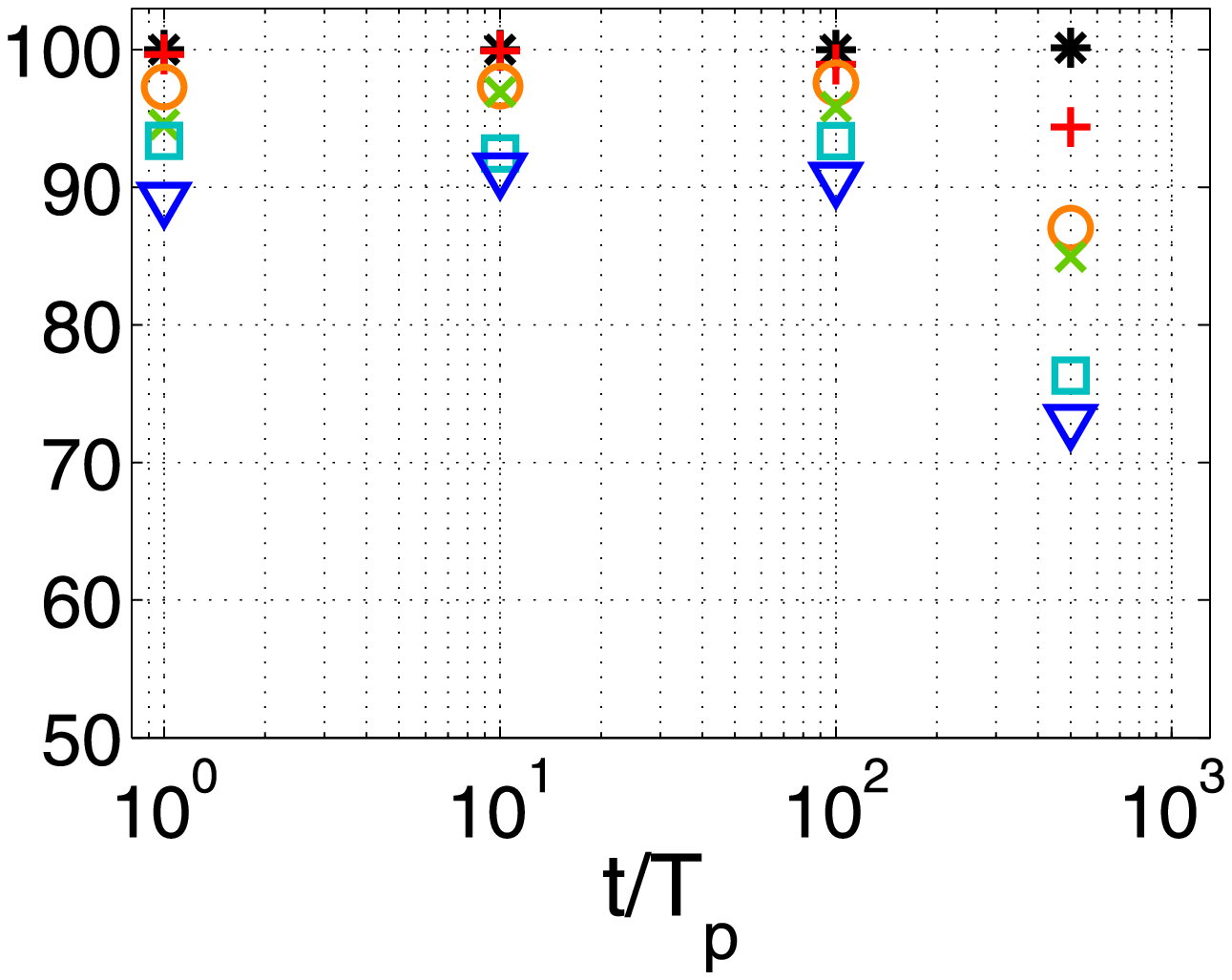}
\put(-130.0,27){(c)}    \put(-15.0,27){(d)}    
\caption{Effect of nonlinearities on the prediction of an oceanic rogue wave in a sea state five. Figures a-d respectively show M=1 (Linear simulation),2,3, and 4 (second, third and fourth order). Markers/colors correspond to different initial perturbations ($E_r$). For example, black star in fig.a shows that the linear model (M=1), even with a zero initial uncertainty ($E_r$=0\%), results in $\sim$5\% error (i.e. $H_{ps}/H_{rs}\sim$0.95) in the prediction of the height of the rogue wave which is only $t=T_p$ ahead. From figures c,d it is seen that convergence is achieved for M=3. Standard error of all data points are less than 2\%. Other simulations parameters are the same as in fig.1.}\label{afoto}
\end{figure}

To highlight the significance of nonlinearities on the prediction and to also provide a convergence test for our scheme, we compare predictions initialized by $\eta_p,\phi_p$, for the rogue wave case presented in figure 1a (sea state five), with $E_r$=0\%, 5\%, 10\%, 15\%, 20\% and 25\%, and by taking different orders of nonlinearity (M) into account (figure 2). To obtain a statistical average of the effect of uncertainty, each presented case (i.e. each marker in figure 2) is the average result of 19 simulations each initiated with an independent set of random perturbations\footnote{The number 19 is chosen to meet error requirements as well as for an efficient use of computational resources at hand.}. All studied cases resulted in standard errors of less than 2\%. 
In each panel of figure 2a-d we also consider four lead times of $t_r=nT_p$, $n=1,10,100,500$.

Linear model (M=1, fig. 2a) under-predicts the height of the rogue wave even with $E_r$=0\% and n=1 (i.e. zero initial disturbance and within $T_p$, i.e. one period of occurrence). For M=2, although for n=1 predictions are acceptable, but for later times results are very much far from converged results of M=3,4 (fig. 2c,d). A good convergence is observed for M=3 (fig. 2c). The general behavior observed in figure 2c,d is qualitatively the same for all cases we investigated.

To see if there is a quantitative trend in the predictability of oceanic rogue waves, we have performed extensive numerical experiments on $\sim O(100)$ rogue waves (documented in our database) for each of the sea states four, five and six. Results for $t_r=500T_p$ are shown in figures 3a,b and c respectively for sea states 4,5 and 6. Each marker is again an average of 19 simulations with standard error of less than 2\%
\footnote{The time $t_r=500T_p$ corresponds to an order of $\sim$1 hour for ocean applications and is chosen because features of results and particularly effects of nonlinearity are more highlighted.}.

Error in the prediction of the height of rogue waves, as suggested by figure 3, is a function of height of the rogue wave, sea state and, of course, degree of uncertainty. For all three sea states, the error in the prediction is higher for higher amplitude rogue waves. The rate of increase in the error is larger for sea state five (fig. 3b) compared with sea state four (fig. 3a). This is due to the higher nonlinearity of the ambient waves in the sea state five that further amplifies disturbances in the evolution equation. Prediction error, however, does not change further from sea state five to sea state six (fig. 3c) that implies an asymptotic saturation of how much the nonlinearity can contribute to the amplification of error over time.

We define predictability horizon for oceanic rogue waves at $H_{ps}/H_{rs}$=0.9, i.e. when rogue wave height prediction can be made with 10\% accuracy%
\footnote{The term ``horizon of predictability'' is used extensively, though not exclusively\cite{Lopez2011a}, in the context of chaotic dynamical systems. We would like to emphasis that, although water surface may undergo chaotic motion in cases \cite{alam2009b}, the use of the term \textit{predictability horizon} here is not based on such behavior, but merely nonlinear amplification of noise by nonlinearity. Whether water surface undergoes chaotic behavior, and if so under what condition(s), is an interesting subject of research but requires a separate investigation.}%
. This definition is based on the fact that 10\% higher $H_{rs}$ corresponds to return period of an order of magnitude longer (in years)\cite{Leonard2013}. Based on this definition, a rogue wave in a sea state four is predictable 500$T_P$ ahead of occurrence if the uncertainty in the sea state measurement at the current time is less than 20\%. To achieve this level of predictability in a sea state 5 and 6, the uncertainty has to be less that $\sim$5\%.

\begin{figure*}
\centering
\includegraphics[height=1.8in]{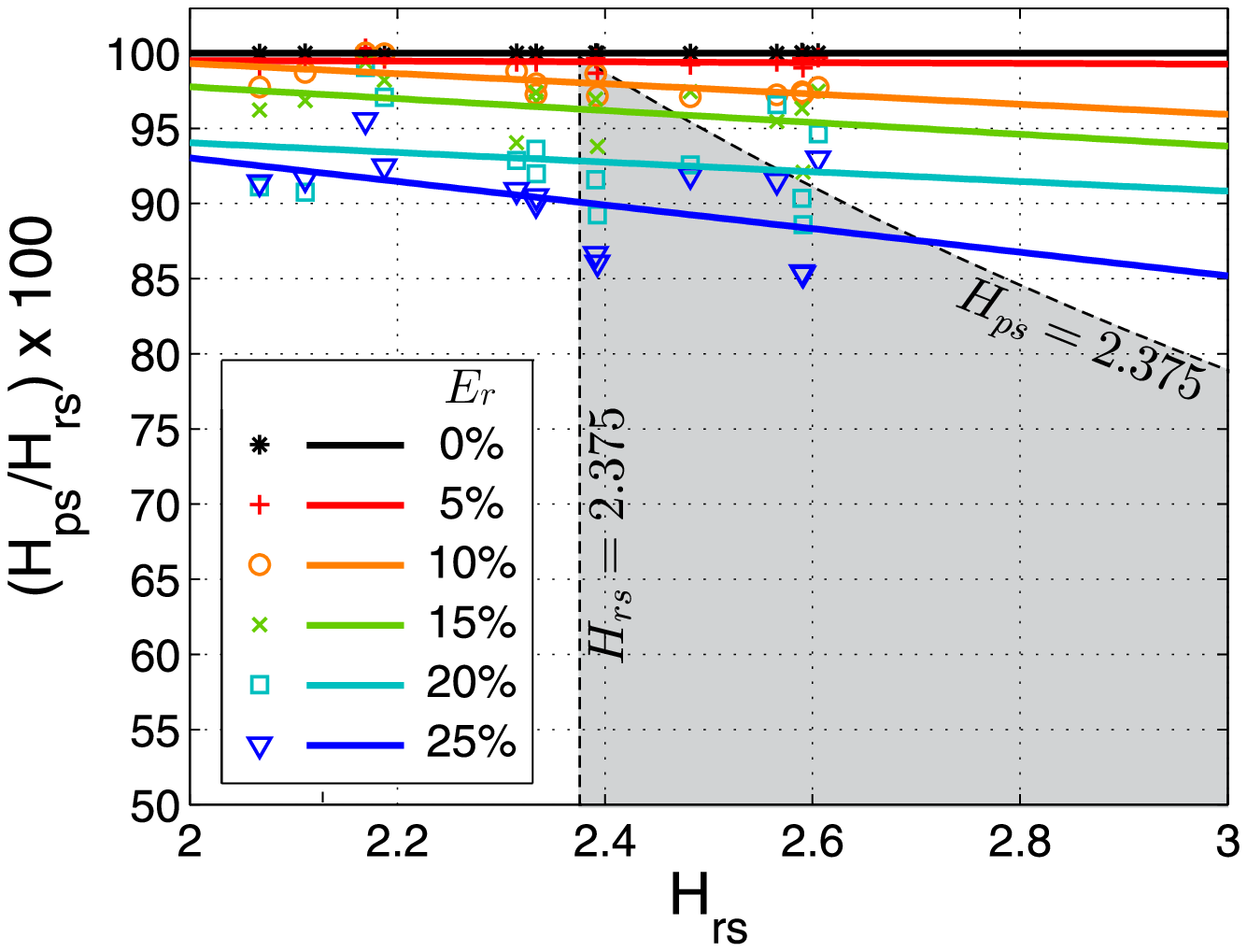}\hspace{0.10cm}
\includegraphics[height=1.8in]{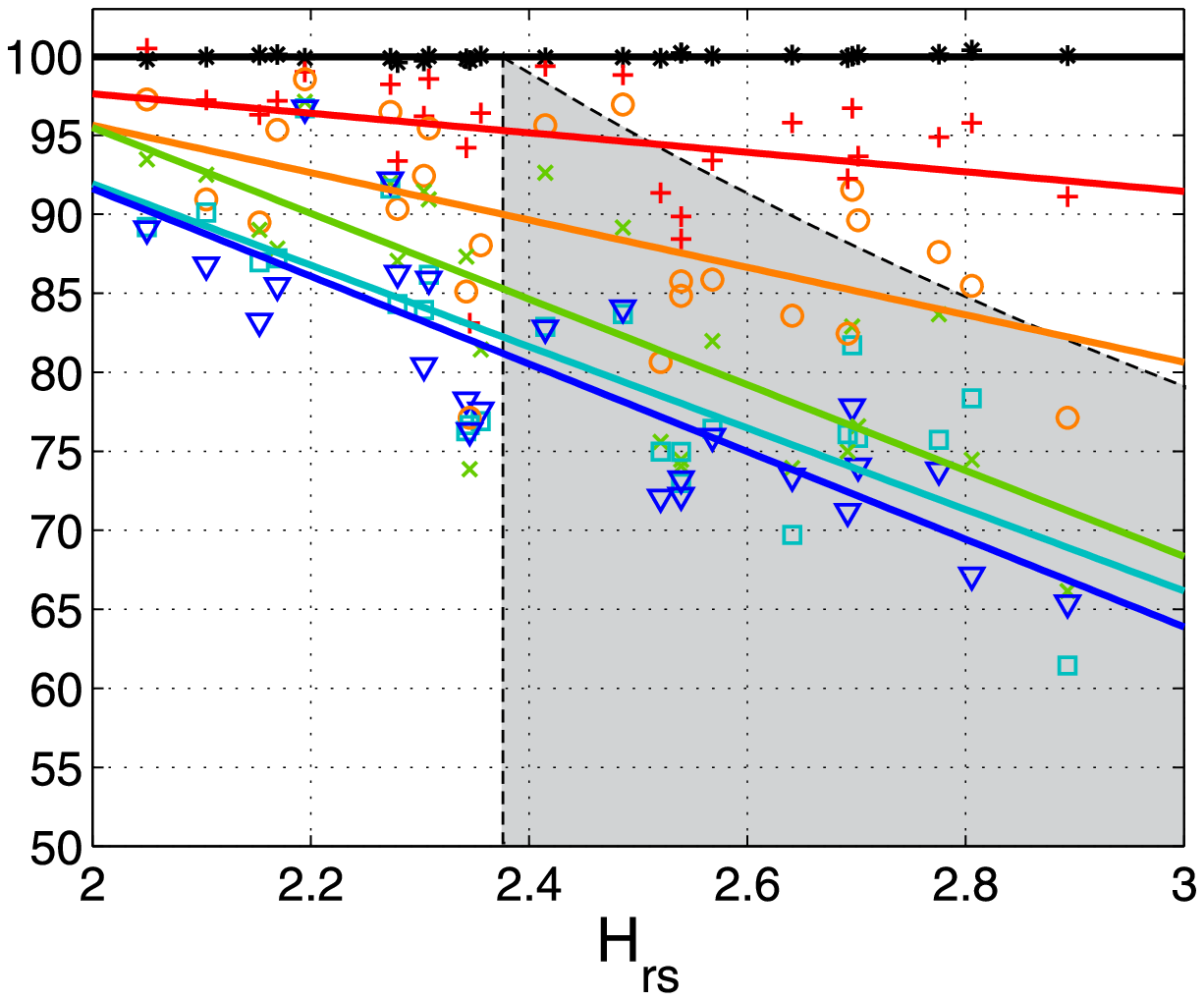}\hspace{0.10cm}
\includegraphics[height=1.8in]{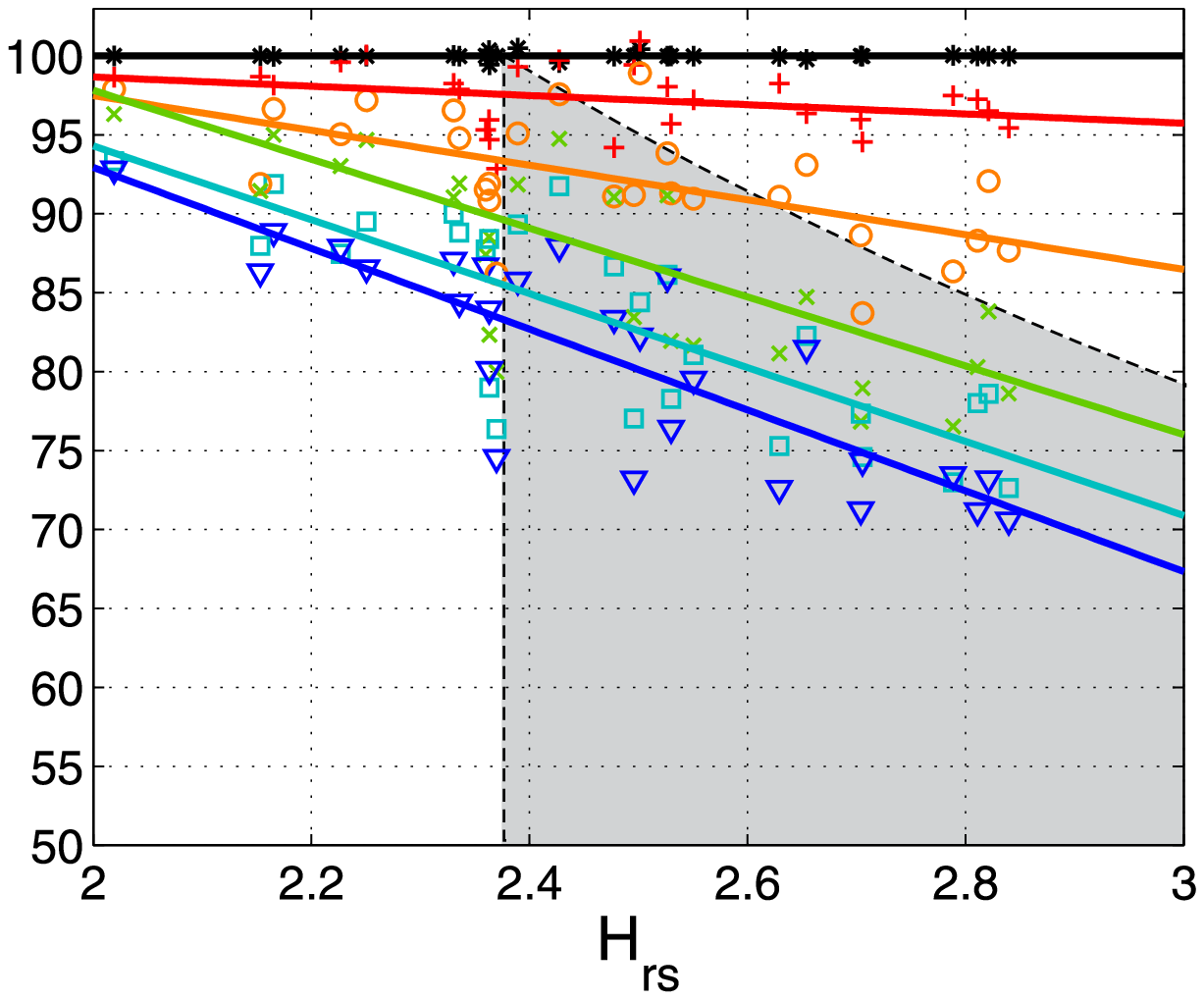}
\put(-345.0,25){(a)} \put(-180,25){(b)} \put(-15,25){(c)}\\
\caption{Predictability of oceanic rogue waves as a function of rogue wave's height and uncertainty in the initial condition ($t_r=$500$T_p$) for respectively sea states four, five and six. For nonzero perturbations error in prediction is larger if amplitude of the rogue wave is larger, and if the sea state is higher (c.f. figs a,b), but reaches an asymptotic saturation at the sea state 6 (c.f. figs. b,c). Shaded areas are particularly important areas for practical applications because actual rogue waves with $H_{rs}>2.375$ (on the right of the vertical dashed line) are stronger than 10,000 year design standards, but for markers that fall inside the shaded area the prediction incorrectly underestimates the amplitude to be safe. Other simulations parameters are the same as in fig.1.}\label{fig2}
\end{figure*}

In practice, many offshore structures are designed today for the extreme waves of return period 10,000 years. It is crucial for these structures to know if a rogue wave with a height greater than the design value is to occur at their location. If this knowledge is in hand several precautionary procedures can be carried out to minimize the damage and the potential life loss (such procedures include, for instance, shut down or relocation). Therefore a critical question is if a reliable prediction can be made. Norwegian offshore standard NORSOK \cite{NORSOKN003} suggests that in the absence of more detailed information an extreme wave of $H_{10000}/H_s$=2.375 has an annual probability of occurrence of less than 10$^{-4}$. Therefore in figures 3a-d, waves with $H_{rs}>$2.375 are larger than $H_{10000}$. Markers that fall inside the gray area show predictions of rogue waves that are in fact larger than $H_{10000}$ (i.e. dangerous to the structure) but are predicted to be smaller than $H_{10000}$ (i.e. falsely predicted to be safe waves). Figures 3b,c show that with more than 10\% uncertaintly in the estimation of wave components of ocean states five and six, rogue waves higher than $H_{10000}$ cannot be predicted sooner than 500$T_p$ ahead of time.

The spectrum considered here is (relatively) broad. It is known, however, that narrower spectra are more amenable to instabilities, and therefore it is expected that the predictability is weaker for a narrower spectrum sea state. The evolution of sea states seven and beyond involves very steep waves, wave breaking and stronger effects of viscous dissipation. Effects of wave-breaking and viscosity are expected to lower the chance of occurrence of oceanic rogue waves and hence positively contribute to the predictability horizon. Both effects can be incorporated into the spectral scheme used here by the means of semi-empirical terms.

To summarize, we defined a quantitative predictability horizon and calculated a statistically-converged predictability time scale for oceanic rogue waves. This horizon is shorter in higher sea states and if the amplitude of the actual anticipated rogue wave is higher. Nonlinearity and nonlinear interactions are the major players behind the amplification of the initial uncertainty, and affect the prediction to the extent that all major features of an upcoming rogue wave may be completely lost.

Sensitivity of predictions of analytical model-equations such as Nonlinear Schr\"{o}dinger\cite{Akhmediev2009,Akhmediev2009a,Chabchoub2011,Akhmediev2010}, its analytical recursive solutions (e.g. Peregrine solitons\cite{Peregrine1983,Shrira2009,Akhmediev2011a} and Akhmediev breather\cite{Mahnke2012,VishnuPriya2013}), and specific growth mechanisms such as Benjamin-Feir instability\cite{Benjamin1967,Xiao2013,Lake1977} (particularly in broadband) to the initial perturbations may provide analytical predictability horizon and worth investigation. Another important and immediate follow up  question is the predictability horizon in three-dimension and how presented results here will be affected. Particularly since Benjamin-Feir instability is less determinant in three-dimension, care must be taken in the investigation and extension of current results to short-crested seas. From a more practical perspective, predictability is also affected by the performance of wave field reconstruction techniques, length of the measurement records, and bandwidth of underlying wave-field.

Rogue waves may appear not only in surface gravity wave systems but also in optical systems \cite{Solli2007}, capillary waves \cite{Shats2010}, and plasma physics \cite{Moslem2011}, where techniques  developed here and results obtained may have similar implications.

Computational resource for this research was provided partially by NERSC (DOE).

\bibliography{536_Rogue_Waves}

\end{document}